\documentclass[english,aps,prb,twocolumn]{revtex4}
\usepackage[T1]{fontenc}
\usepackage[latin9]{inputenc}
\usepackage{xcolor}
\usepackage{pdfcolmk}
\usepackage{textcomp}
\usepackage{amsthm}
\usepackage{amsmath}
\usepackage{amssymb}
\usepackage{graphicx}
\PassOptionsToPackage{normalem}{ulem}
\usepackage{ulem}

\makeatletter

\providecolor{lyxadded}{rgb}{0,0,1}
\providecolor{lyxdeleted}{rgb}{1,0,0}

\@ifundefined{textcolor}{}
{%
 \definecolor{BLACK}{gray}{0}
 \definecolor{WHITE}{gray}{1}
 \definecolor{RED}{rgb}{1,0,0}
 \definecolor{GREEN}{rgb}{0,1,0}
 \definecolor{BLUE}{rgb}{0,0,1}
 \definecolor{CYAN}{cmyk}{1,0,0,0}
 \definecolor{MAGENTA}{cmyk}{0,1,0,0}
 \definecolor{YELLOW}{cmyk}{0,0,1,0}
}

\makeatother

\usepackage{babel}
\begin{document}

\title{Time-Optimal Universal Control of Two-Level Systems under Strong
Driving}

\author{C. Avinadav, R. Fischer, P. London, and D. Gershoni}

\affiliation{Department of Physics, Technion, Israel Institute of Technology,
Haifa 32000, Israel}
\begin{abstract}
We report on a theoretical and experimental study of time-optimal
construction of arbitrary single-qubit rotations under a single strong
driving field of finite amplitude. Using radiation-dressed states
of nitrogen vacancy centers in diamond we realize a strongly-driven
two-level system, with driving frequencies four times larger than
its precession frequency. We implement time optimal universal rotations
on this system, characterize their performance using quantum process
tomography, and demonstrate a dual-axis multi-pulse control sequence
where the qubit is rotated on time scales faster than its precession
period. Our results pave the way for applying fast qubit control and
high-density pulse schemes in the fields of quantum information processing
and quantum metrology.
\end{abstract}
\maketitle

\section{Introduction}

Two-level systems are the prototypical realization of a quantum bit
(qubit), and as such, coherent control of their state is a key element
in novel quantum devices and applications. Two important measures
of qubit control techniques are the \emph{time} it takes to complete
a desired rotation, and the ability to perform \emph{arbitrary} state
rotations within a given realization. Reduced manipulation times allow
for an increased number of operations to be performed within the system
coherence time, a critical requisite of quantum information processing
\cite{DiVincenzo2000} and quantum sensing \cite{deLange2011}. Similarly,
the ability to perform universal single-qubit rotations \cite{Foletti2009,Kodriano2012}
can reduce the complexity of quantum computing algorithms as opposed
to utilizing only a minimal set of single-qubit gates \cite{Galindo2002,Dawson2006}.
Universal rotations are also useful in pulsed quantum sensing schemes
\cite{Zhao2012}, as they allow for systematic pulse error compensation
using composite pulses \cite{Levitt1979473,Vandersypen2005} or multi-axis
decoupling techniques \cite{Gullion1990,deLange2010,Wang2012}. 

Generally, the rotation time of a qubit\textquoteright{}s state depends
on the strength of the fields applied to it. Obtaining shorter manipulation
times therefore requires stronger driving fields, eventually leading
to the strong driving regime. In this regime the qubit is driven by
an external field whose coupling energy is comparable to, or exceeds,
the energy level splitting of the qubit itself. The traditional method
of applying an oscillatory field at the qubit\textquoteright{}s resonance
frequency results in complex dynamics under strong driving, due to
the counter-rotating term of the oscillating field. This term can
be neglected for weak driving according to the rotating wave approximation
\cite{Jaynes,Frasca2003193}, but for strong fields it plays a crucial
role in the system dynamics \cite{Fuchs2009,Chiorescu2004}. Therefore,
different control schemes have been considered for this regime, including
usage of ancillary energy levels \cite{Kodriano2012}, two independent
orthogonal fields \cite{Shim2014,London2014}, or single-field non-harmonic
pulses.

Considering the approach of non-harmonic pulse sequences it was shown,
with tools of optimal control theory, that the fastest way to steer
a qubit on the Bloch sphere from one state to another, using a single
driving field of finite amplitude $\left|B\left(t\right)\right|\le B_{\textrm{max}}$,
is a \emph{bang-bang} control \cite{Boscain2005,Boscain2006}, i.e.
rectangular pulses alternating between the extremal values of the
driving field $\pm B_{\textrm{max}}$. This approach was recently
applied also to the Landau-Zener problem, for finding the \textquotedblleft{}quantum
speed limit\textquotedblright{}, the minimal time required to transfer
a system between any two states \cite{Hegerfeldt2013}. However, these
studies considered only the problem of \emph{steering the system state}
between two points on the Bloch sphere, rather than \emph{generating
a prescribed rotation operator}. Thus, for example, these results
cannot be applied for achieving $\pi$-flips around an \emph{arbitrary}
rotation axis. The challenge of generating \emph{any} desired unitary
operator was previously studied only for weak driving \cite{Garon2013}
or under the assumption of infinitely strong driving fields \cite{Khaneja2001}.

In this work we present a theoretical and experimental study of time-optimal
universal qubit rotations in the strong driving regime. We derive
the necessary conditions that must be satisfied by a pulse sequence
in order to be time-optimal, and through numerical optimization design
pulses for the important cases of $\pi/2$ and $\pi$-rotations around
arbitrary axes in the Bloch sphere\textquoteright{}s equatorial plane.
Then, we experimentally apply these control sequences on radiation-dressed
states of electron spins in nitrogen-vacancy (NV) centers in diamond.
This approach enables us to realize a strongly-driven two-level system
with excellent controllability and superior coherence properties with
respect to bare NV spin. Finally we use this system to apply a dual-axis
multi-pulse sequence with an unprecedented inter-pulse delay of two
spin precession periods.

\section{Time-optimal synthesis of universal rotations}

We consider a general two-level system, or a manifold of a more complex
system, driven by a single external field. It is described by the
time-dependent Hamiltonian
\begin{equation}
H\left(t\right)=\frac{1}{2}\hbar\omega_{1}\sigma_{z}+\frac{1}{2}\hbar\Omega\left(t\right)\sigma_{x},\label{eq:GeneralHamiltonian}
\end{equation}
where $\omega_{1}$ is the energy level splitting, or the spin precession
frequency, $\Omega\left(t\right)$ is the applied driving field bounded
by $\left|\Omega\left(t\right)\right|\le\Omega_{\textrm{max}}$, and
$\sigma_{i}$ are the Pauli matrices, where $\hat{x}$ is the direction
of the applied external field. The time evolution operator $U\left(t\right)$
represents the rotation induced by the control sequence $\Omega\left(t\right)$,
and evolves in time according to
\begin{equation}
i\hbar\dot{U}\left(t\right)=H\left(t\right)U\left(t\right),\label{eq:Schrodinger}
\end{equation}
with $U\left(0\right)=I_{2\times2}$. In this formalism, the problem
of generating a time-optimal rotation can be regarded as that of steering
the operator $U\left(t\right)$, which lies in SU(2), in minimal time
onto the desired rotation $U_{\textrm{final}}$. We solve this problem
in a two-step process of \emph{reduction} and \emph{selection}. First,
we apply Pontryagin\textquoteright{}s minimum principle (PMP) \cite{Pontryagin1987,Ross2009}
which gives the necessary conditions that a control $\Omega\left(t\right)$
must satisfy in order to be optimal, thus reducing substantially the
number of optimal control candidates. Second, we select from these
candidates the control sequences which satisfy the problem, i.e. generate
the desired rotations, and choose the one which does so in minimal
time. We outline these steps below and present the resulting time-optimal
controls.

First, using Pontryagin's minimum principle we prove in the Appendix
that time-optimal control sequences consist only of \emph{bang periods},
in which $\Omega\left(t\right)\equiv\pm\Omega_{\textrm{max}}$ and
the qubit rotates about an axis $\omega_{1}\hat{z}\pm\Omega_{\textrm{max}}\hat{x}$,
and \emph{drift periods}, where $\Omega\left(t\right)\equiv0$ and
the qubit simply precesses around the z-axis. Remarkably, any desired
single-qubit operator may be constructed in this way. This result
substantially reduces the number of candidate sequences for optimal
control. Finding the optimal sequence therefore amounts to selecting
the correct number, ordering and length of the bang and drift periods.
While this is still a challenging problem to solve analytically, it
can be approached by numerical optimization on an $n$-dimensional
space, where $n$ is the assumed number of bang or drift periods and
the optimization variables represent their durations. These durations
are bounded, as the dynamics under bang or drift controls are periodic,
with $T_{b}=2\pi/\sqrt{\omega_{1}^{2}+\Omega_{\textrm{max}}^{2}}$
or $T_{d}=2\pi/\omega_{1}$, respectively. Since each pulse may either
be a positive bang, a negative bang or a drift, the optimization is
repeated $3\times2^{n-1}$ times to explore all unique pulse sequences%
\footnote{An $n$-pulse sequence that contains consecutive pulses of the same
type can be considered as an $\left(n-1\right)$-pulse sequence, and
needn\textquoteright{}t be calculated again. Thus in constructing
unique sequences we have 3 options for the initial pulse type, but
only 2 options for the remaining $n-1$ pulses.%
}. Using this method we verified up $n=12$ to that three pulses or
less are sufficient to construct any rotation operator, in the strong
driving regime $\Omega_{\textrm{max}}\ge\omega_{1}$, in agreement
with the ansatz made in \cite{Hegerfeldt2013}.

\begin{figure}
\includegraphics[bb=0bp 30bp 720bp 500bp,clip,width=1\columnwidth]{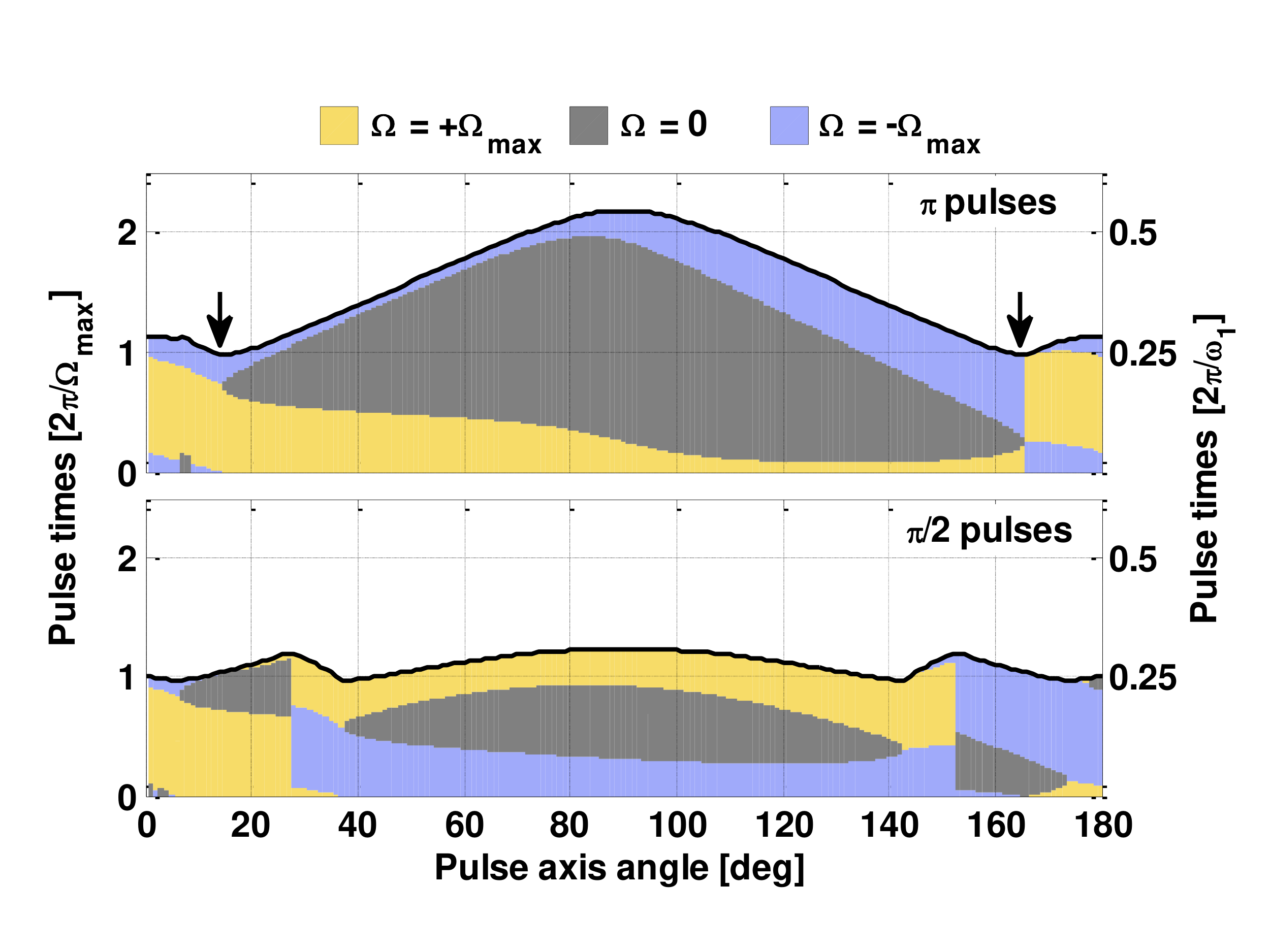}

\caption{(color online) Time-optimal pulse sequences for generating $\pi/2$
and $\pi$-rotations (top and bottom panels respectively), for $\Omega_{\textrm{max}}=4\omega_{1}$.
Colors represent different types of pulses: positive/negative bangs
and drift. Black arrows mark the globally-optimal $\pi$-rotations,
corresponding to pure bang-bang controls.\label{fig:OptimalPulseMaps}}
\end{figure}

We then set to calculate the pulse sequences that generate time-optimal
$\pi/2$ and $\pi$-rotations around different axes in the x-y plane
{[}Fig. \ref{fig:OptimalPulseMaps}{]}. In the case of $\pi$-pulses
we note the following: (a) for $\Omega_{\textrm{max}}\gg\omega_{1}$
the duration of the \emph{longest} control sequence approaches $\pi/\omega_{1}$,
corresponding to half a precession cycle of the spin inserted between
two $\delta$-like $\pi/2$-rotations; (b) the \emph{shortest} control
sequence consists of two opposite bang periods and lasts exactly $2\pi/\sqrt{\omega_{1}^{2}+\Omega_{\textrm{max}}^{2}}$,
independent of the ratio of $\Omega_{\textrm{max}}$ to $\omega_{1}$.
These two results are consistent with previous studies of infinite-amplitude
fields \cite{Khaneja2001} and shortest possible spin flips \cite{Boscain2005,Boscain2006}.

\section{Experimental system}

We use radiation-dressed states of the electron spin of nitrogen-vacancy
(NV) centers in diamond \cite{Jelezko2006} to realize a strongly-driven
two-level system, and to implement the control sequences described
above. This approach of using dressed states as an effective two-level
system can be applied to many different physical systems, and it offers,
among other advantages which will be explained below, a high degree
of controllability of the system and extended coherence times \cite{Timoney2011}.
The scheme is based on resonant microwave radiation interacting with
the NV center electron spin, thereby creating a dressed two-level
system whose energy level splitting is determined by the coupling
energy of the dressing field. A second magnetic field, orthogonal
to the resonant microwave field, is used to manipulate the state of
the dressed qubit on its Bloch sphere.

\subsection{Radiation-dressed states of NV centers}

\begin{figure}
\includegraphics[bb=0bp 360bp 260bp 540bp,clip,width=1\columnwidth]{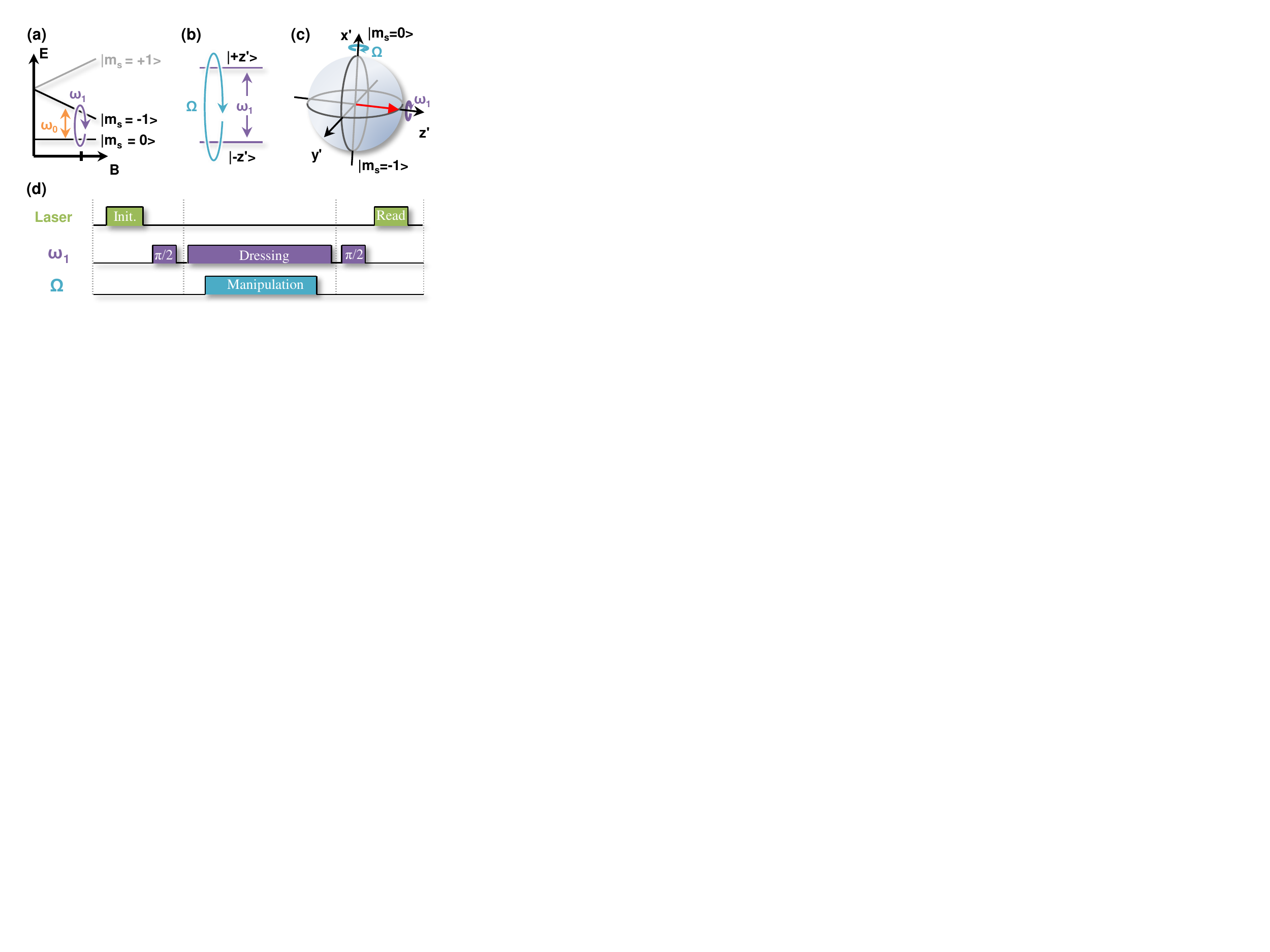}

\caption{(color online) (a) Energy level structure of the NV center ground
state in the presence of axial magnetic field B, which lifts the degeneracy
between the $m_{s}=\pm1$ states. (b) Applying continuous microwave
excitation resonant with the $m_{s}=0,-1$ transition creates new
dressed states with an energy separation $\omega_{1}$, determined
by the Rabi frequency of the dressing field. (c) Bloch sphere representation
of the electron spin in the $m_{s}=0,-1$ subspace. The continuous
driving field $\omega_{1}$ defines the new $z^{\prime}$-axis of
the dressed qubit basis, and the second-order driving $\Omega\left(t\right)$
acts in an orthogonal direction, enabling manipulation of the dressed
states. (d) Experimental sequence: initialization of the dressed spin
by optical pumping and a microwave $\frac{\pi}{2}$-pulse; manipulation
of the dressed spin with the second order field $\Omega\left(t\right)$,
while the microwave field continuously dresses the spin; and readout
of the dressed spin projections by a microwave $\frac{\pi}{2}$-pulse
and optical readout of the bare spin population.\label{fig:DressedStates}}
\end{figure}

The NV center has a spin triplet ($S=1$) ground state {[}Fig. \ref{fig:DressedStates}(a){]}.
It can be optically pumped into the $m_{s}=0$ state with a short
laser pulse, and its population can be measured by spin-dependent
photoluminescence \cite{Manson2006}. A magnetic field of 540 G aligned
to the NV $z$-axis lifts the degeneracy of the $m_{s}=\pm1$ states
and enables selective microwave excitation of the $m_{s}=0,-1$ transition
at $\omega_{0}=\left(2\pi\right)1.36$ GHz. This selectivity allows
us to consider only the $m_{s}=0,-1$ states of the NV center spin,
corresponding to the $\left|+z\right\rangle $ and $\left|-z\right\rangle $
eigenstates of a pseudo spin-$\frac{1}{2}$ system, while the $m_{s}=+1$
state is well out of resonance with any of the applied fields and
does not participate in the system dynamics. 

We write the system Hamiltonian as
\begin{equation}
H\left(t\right)=\frac{1}{2}\hbar\omega_{0}\sigma_{z}+\hbar\omega_{1}\cos\left(\omega_{0}t\right)\sigma_{x}+\frac{1}{2}\hbar\Omega_{\textrm{max}}A\left(t\right)\sigma_{z},
\end{equation}
where the first term corresponds to the unperturbed Hamiltonian of
the pseudo spin-$\frac{1}{2}$, the second term describes the resonant
microwave field applied along the NV center $x$-axis (chosen arbitrarily)
that creates the dressed qubit, and the third term is the second-order
magnetic field applied along the $z$-axis that allows manipulation
of the dressed qubit states. $\omega_{0}$ is the splitting between
the two bare spin eigenstates ($m_{s}=0,-1$), $\omega_{1}$ is the
Rabi frequency of the dressing field, and $\Omega_{\textrm{max}}$
is the maximum amplitude of the second-order field with $A\left(t\right)$
its envelope function. In the interaction picture of $H_{0}=\frac{1}{2}\hbar\omega_{0}\sigma_{z}$
and under the rotating wave approximation (RWA) for $\omega_{1}\ll\omega_{0}$,
the Hamiltonian becomes
\begin{equation}
\tilde{H}\left(t\right)=\frac{1}{2}\hbar\omega_{1}\sigma_{x}+\frac{1}{2}\Omega_{\textrm{max}}A\left(t\right)\sigma_{z}.\label{eq:DressedHamiltonian}
\end{equation}
Upon the rotation $\left(x,y,z\right)\rightarrow\left(z^{\prime},y^{\prime},-x^{\prime}\right)$,
which represents the transformation into the dressed-states basis
{[}Fig. \ref{fig:DressedStates}(b)-(c){]}, this Hamiltonian exactly
manifests the prototypical Hamiltonian of Eq. (\ref{eq:GeneralHamiltonian}).
It describes the dressed qubit, a two-level system with energy level
splitting $\hbar\omega_{1}$, whose eigenstates correspond to the
$\left|+x\right\rangle $ and $\left|-x\right\rangle $ states of
the bare NV electron spin in the rotating frame. The dressed qubit
is driven by a time-dependent transverse field $\Omega_{\textrm{max}}A\left(t\right)$.
Thus, by tuning the ratio between $\Omega_{\textrm{max}}$ and $\omega_{1}$
one can switch between the weak and strong driving regimes, and study
various manipulation techniques using different envelope functions
$A\left(t\right)$. Our experiments were conducted with the parameters
$\omega_{1}=\left(2\pi\right)1.5$ MHz and $\Omega_{\textrm{max}}=\left(2\pi\right)6$
MHz, representing a driving field \emph{four times stronger} than
the dressed spin\textquoteright{}s precession frequency.

Experiments on the dressed qubit consist of three stages: arbitrary
state initialization, pulsed manipulation, and tomographic state readout
{[}Fig. \ref{fig:DressedStates}(d){]}. The dressed qubit is initialized
to $\left|+z^{\prime}\right\rangle $ by optically pumping the spins
into the $\left|+z\right\rangle $ state (corresponding to $m_{s}=0$)
using a short laser pulse, and then rotating them to $\left|+x\right\rangle $
by a $\left(\frac{\pi}{2}\right)_{y}$-rotation induced with a short
pulse of the microwave field. Similarly, to measure the dressed qubit\textquoteright{}s
$z^{\prime}$-projection we apply the same sequence in reverse, i.e.
a $\left(\frac{\pi}{2}\right)_{y}$-pulse followed by a laser pulse
to readout the $m_{s}=0$ population of the bare spins. By changing
the duration and rotation axis of the $\frac{\pi}{2}$ pulses in either
stage we may initialize the dressed spin into different states or
measure different projections of it, thereby enabling arbitrary initialization
and complete state tomography of the dressed qubit.

\subsection{Advantages of using the dressed spin}

Using the radiation-dressed spin in this experimental study of fundamental
control in quantum systems offers several distinct advantages over
the bare spin system:

(a) the dressed spin is protected from spurious magnetic noise via
continuous dynamical decoupling \cite{Chen2006,Cai2012,Xu2012}. In
our setup, we measured a ten-fold improvement in the phase-memory
time $T_{2}^{*}$ from 0.7 to 7 $\mu\textrm{s}$.

(b) The dressed spin accurately manifests the Hamiltonian {[}Eq. \ref{eq:GeneralHamiltonian}{]},
and specifically is insensitive to the alignment of the second-order
driving field $\Omega\left(t\right)$: any $x$ or $y$ components
of this field average to zero in the rotating frame under the RWA
since $\Omega_{\textrm{max}}\ll\omega_{0}$, resulting in pure transverse
driving of the dressed spin and eliminating dynamical energy shifts
that may cause unwanted coupling between its states.

(c) The dressed spin is a better representation of a true two-level
system than the bare system, which may include additional energy states.
This is particularly important for the application of bang-bang control
sequences, which are intrinsically wideband and may therefore cause
unwanted coupling to these additional energy levels \cite{Wang2011}.
With this analogy to a two-level system in mind, and for the sake
of brevity, henceforth we refer to the dressed spin simply as a qubit
or a spin.

\subsection{Harmonic driving of the dressed spin beyond the RWA}

\begin{figure}
\includegraphics[bb=0bp 360bp 250bp 540bp,clip,width=1\columnwidth]{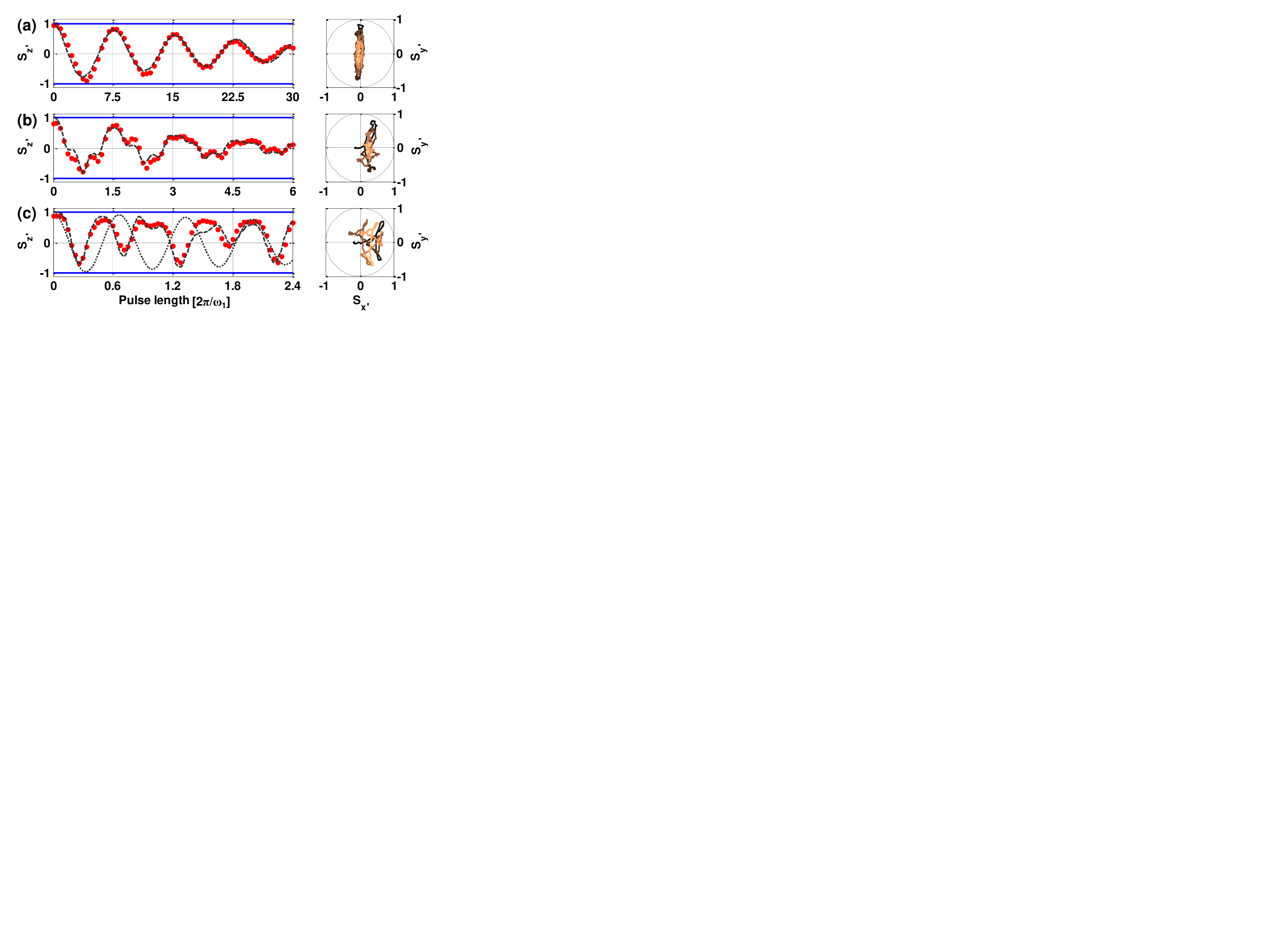}

\caption{(color online) Harmonic driving of the dressed spin at increasing
coupling strengths: (a) $\Omega_{\textrm{max}}/\omega_{1}=0.16$,
(b) $\Omega_{\textrm{max}}/\omega_{1}=0.67$, and (c) $\Omega_{\textrm{max}}/\omega_{1}=1.5$.
Left panels show the $z^{\prime}$ component of the spin (dots: experiment,
thick dashed line: simulation, thin dotted line: expected Rabi oscillations
without the counter-rotating term) and right panels show the spin
trajectory in the $x^{\prime}$-$y^{\prime}$ plane (measurements
only, time flows from dark to bright colors). An exponential decaying
envelope was fitted to the measurements and applied to the simulation
results for comparison purposes.\label{fig:StrongHarmonicDriving}}
\end{figure}

Here we demonstrate the dressed spin dynamics under resonant harmonic
excitation, i.e. $\Omega\left(t\right)=\Omega_{\textrm{max}}\cos\left(\omega_{1}t\right)$.
In a frame rotating at $\omega_{1}$ this field is rewritten as the
sum of two terms, $\Omega\left(t\right)=\Omega_{\textrm{max}}\left[1+\exp\left(2i\omega_{1}t\right)\right]/2$.
The first term is fixed in the rotating frame, while the second rotates
at twice the precession frequency $\omega_{1}$. At the weak driving
regime, $\Omega_{\textrm{max}}\ll\omega_{1}$, the second term averages
to zero \textendash{} this is the rotating wave approximation \textendash{}
and the resulting spin dynamics are the familiar Rabi oscillations,
occurring at a frequency $\Omega_{\textrm{max}}$ {[}Fig. \ref{fig:StrongHarmonicDriving}(a){]}\cite{Loretz2013}.

However if we attempt to shorten the spin manipulation time by increasing
the field amplitude, the counter-rotating term becomes non-negligible
and the dynamics differ significantly from the expected result in
both the $z^{\prime}$-projection of the spin and its $x^{\prime}$-$y^{\prime}$
components {[}Fig. \ref{fig:StrongHarmonicDriving}(b)-(c){]}. These
results show a clear signature of a strongly-driven two-level system.
It is important to note that while the dressed spin does rotate on
time scales much shorter than its precession period in Fig. \ref{fig:StrongHarmonicDriving}(c)
for $\Omega_{\textrm{max}}/\omega_{1}=1.5$, the first \textquotedblleft{}dip\textquotedblright{}
of the oscillations that occurs at $t=0.25\times\left(2\pi/\omega_{1}\right)$
does not correspond to an actual spin flip, i.e. the spin does not
reach the south pole of the Bloch sphere. It is only on the third
dip, at $t=1.25\times\left(2\pi/\omega_{1}\right)$, that the spin
actually reaches this state, demonstrating the inefficiency of the
harmonic driving method at such strong fields.

\section{Demonstration of time-optimal controls and their application}

\subsection{Characterization of time-optimal $\boldsymbol{\pi}$-pulses}

\begin{figure}
\includegraphics[bb=0bp 350bp 270bp 540bp,clip,width=1\columnwidth]{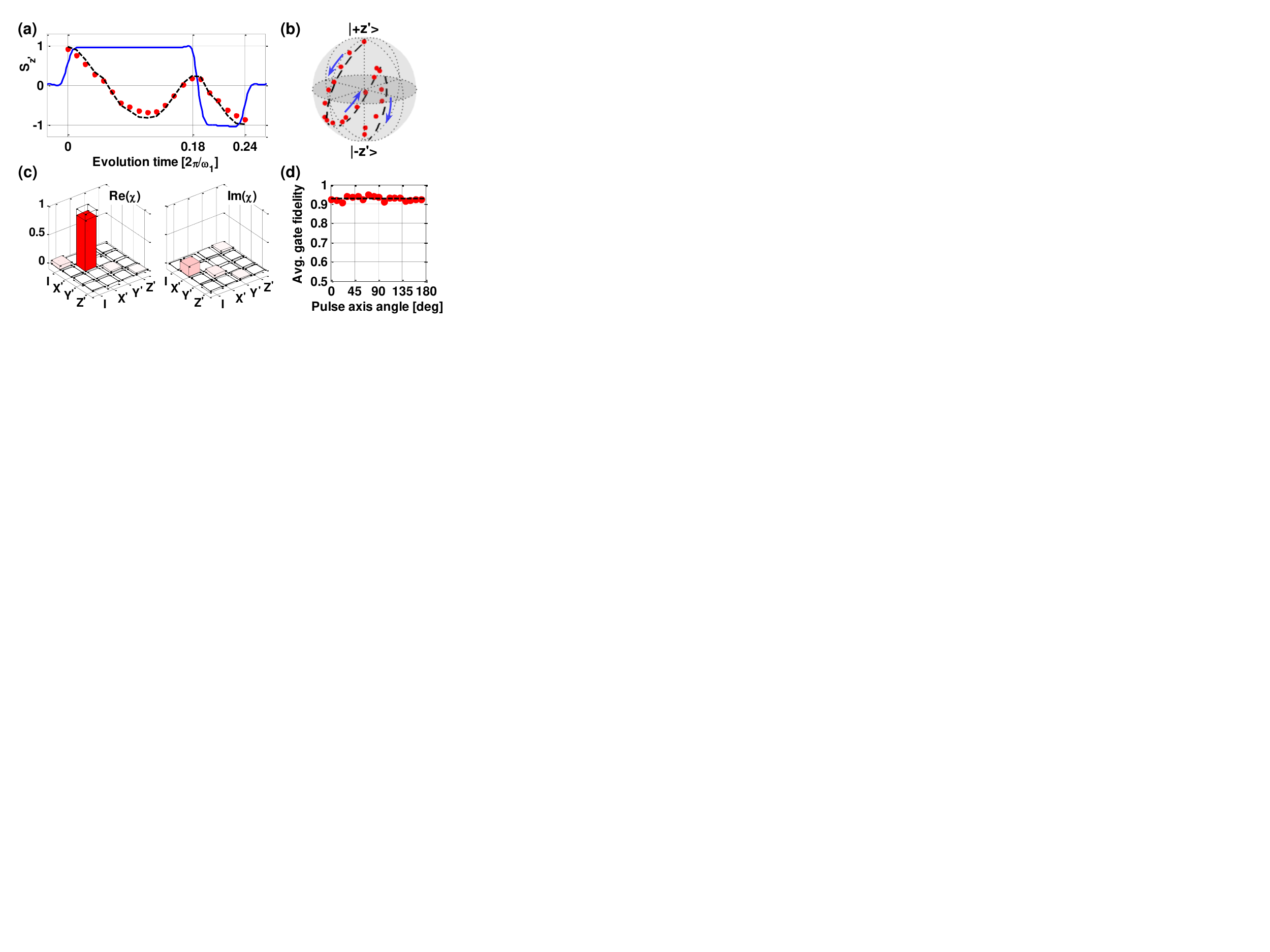}

\caption{(color online) (a) and (b): Measurement of the spin $z^{\prime}$-component
and its full state trajectory, respectively, under a bang-bang sequence
(red dots: experiment, black dashed line: simulation, blue solid line:
pulse sequence in units of $\Omega_{\textrm{max}}$. Arrows denote
the flow of time). (c) Real and imaginary parts of the measured process
matrix for $\pi_{x^{\prime}}$-pulse (empty bars show the ideal matrix).
(d) Measured gate fidelities of $\pi$-pulses around different axes
in the $x^{\prime}$-$y^{\prime}$ plane. Average fidelity obtained
was 0.93\textpm{}0.01.\label{fig:PiPulseCharacterization}}
\end{figure}

We implemented the driving scheme described in Section II on an ensemble
of NV centers using the dressed spin as the prototype two-level system,
and measured the performance of the designed time-optimal $\pi$-pulses.
We first illustrate the spin dynamics under such pulses in detail.
For a bang-bang control sequence bounded by $\Omega_{\textrm{max}}=4\omega_{1}$,
the qubit is rotated on time-scales faster than its precession frequency
$\omega_{1}$ {[}Fig. \ref{fig:PiPulseCharacterization}(a){]}. The
measured state trajectory on the Bloch sphere under this sequence
{[}Fig. \ref{fig:PiPulseCharacterization}(b){]} fits the predicted
behavior well: the first bang rotates the qubit from the north pole
$\left|+z^{\prime}\right\rangle $ to the equator, failing to pass
through the desired south pole $\left|-z^{\prime}\right\rangle $
due to the non-vanishing free precession $\omega_{1}$. The second
bang, of opposite sign, compensates for this effect and completes
the $\pi$-pulse by rotating the qubit exactly to $\left|-z^{\prime}\right\rangle $.

A complete characterization of universal spin-flip sequences was carried
out using quantum process tomography \cite{Nielsen2000}. Figure \ref{fig:PiPulseCharacterization}(c)
shows the measured process matrix for a $\pi$-pulse around the x-axis,
with an average gate fidelity%
\footnote{The average gate fidelity is given by $F_{g}=\left(d\cdot F_{\chi}+1\right)/\left(d+1\right)$,
where $d=2$ is the system dimension and $F_{\chi}=\textrm{Tr}\left[\chi_{\textrm{meas}}\chi_{\textrm{ideal}}\right]$
is the process matrix fidelity; see M. A. Nielsen, Phys. Lett. A 303,
249 (2002).%
} of 0.92. $\pi$-flips around different axes in the x-y plane were
also characterized {[}Fig. \ref{fig:PiPulseCharacterization}(d){]}.
We measure an average gate fidelity of 0.93\textpm{}0.01. The fidelity
is limited by our technical ability to deliver ideal bang-bang pulses,
and does not represent a fundamental limit of our spin control technique.
These results demonstrate our ability to perform universal qubit rotations
at time-scales much shorter than the qubit's precession period. Achieving
similar controllability using traditional harmonic driving requires
at least an order of magnitude \emph{slower} dynamics than the precession
period in order to satisfy the RWA and to maintain high fidelity.

\subsection{Application of time-optimal control in quantum protocols}

In addition to quantum information processing, fast arbitrary-axis
rotations also play a key role in ac magnetometry sequences, aimed
at detecting classical sources emitting ac magnetic fields \cite{Laraoui2010,deLange2011}
or the fluctuating fields of nearby spins \cite{Staudacher2013,Shi2013}.
One such sensing scheme is based on repeatedly applying $\pi$-pulses
to the sensing spin at regular intervals \cite{Taylor2008,Staudacher2013},
similar to pulsed dynamical decoupling sequences \cite{Viola1999,Uhrig2007}.
This acts as a lock-in measurement, significantly increasing the sensitivity
of the spin to ac magnetic fields at frequencies matching the inter-pulse
delay. Keeping this delay fixed and increasing the number of $\pi$-pulses
improves the sensitivity and narrows the frequency response function.
However two deteriorating mechanisms compete with this improvement:
(a) the total sequence time increases linearly with the number of
pulses while the coherence time improves only sub-linearly \cite{deLange2010,Bar-Gill2012},
exposing the spin to decoherence; (b) errors in the $\pi$-pulses
accumulate and decrease the overall signal fidelity.

\begin{figure}
\includegraphics[bb=0bp 340bp 250bp 540bp,clip,width=1\columnwidth]{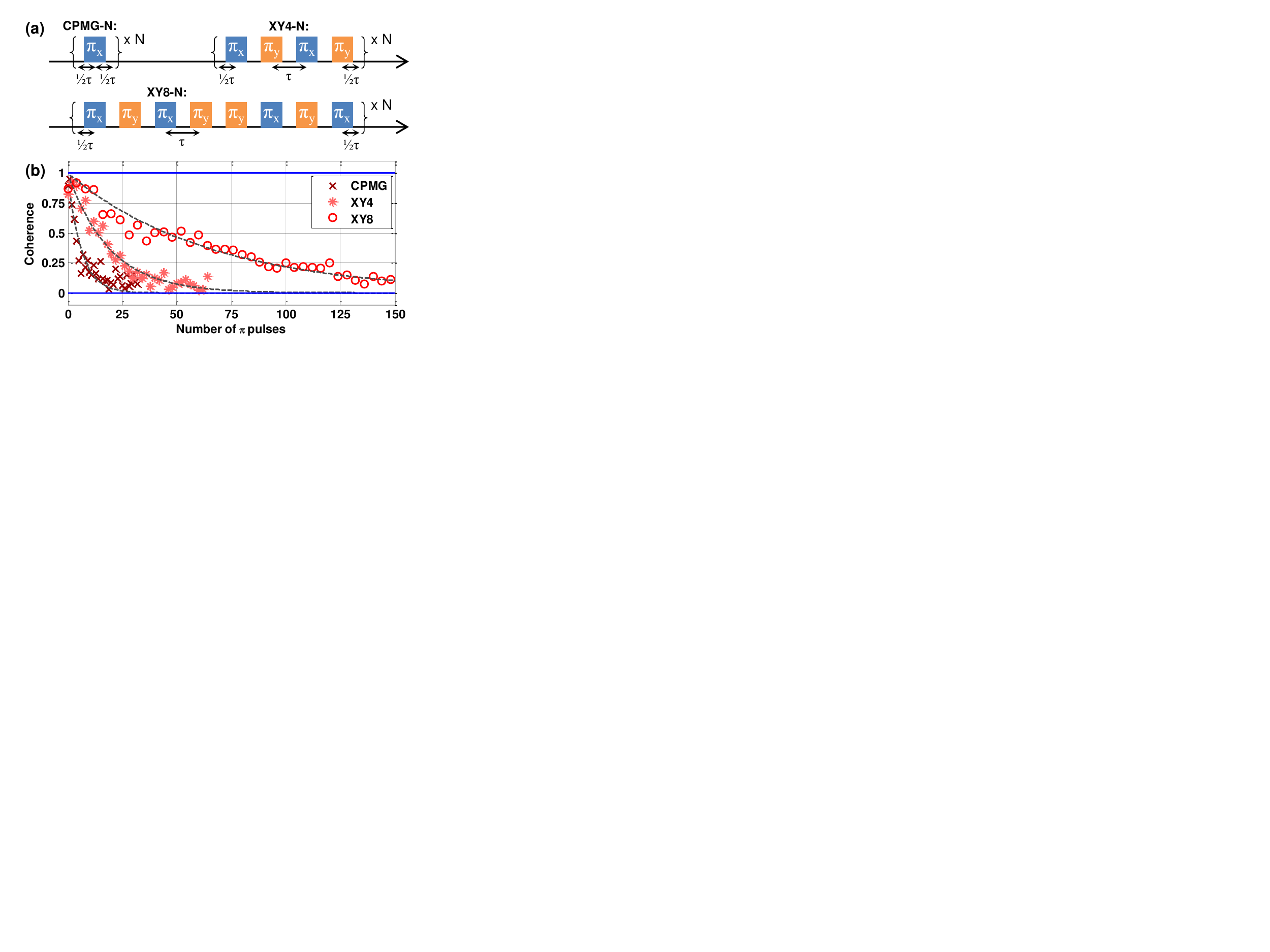}

\caption{(color online) Demonstration of ac magnetometry sequences using two-axis
control in the strong driving regime. (a) CPMG-N: a fixed-axis $\pi_{x}$
pulse is applied $N$ times at regular intervals; XY4-N and XY8-N:
the sequences $\pi_{x}-\pi_{y}-\pi_{x}-\pi_{y}$ and $\pi_{x}-\pi_{y}-\pi_{x}-\pi_{y}-\pi_{y}-\pi_{x}-\pi_{y}-\pi_{x}$,
respectively, are applied $N$ times. At the beginning and end of
each sequence a $\left(\pi/2\right)_{x}$ pulse is applied. (b) Measured
coherence as a function of the total number of $\pi$ pulses, under
different decoupling sequences (symbols: experiment, dash lines: exponential
decay fits). In all cases the inter-pulse delay was two precession
periods ($4\pi/\omega_{1}$).\label{fig:DynamicalDecoupling}}
\end{figure}

The effect of pulse error accumulation can be mitigated by implementing
multi-axis control over the spin and by symmetrizing the pulse sequence
\cite{Gullion1990,deLange2010,Wang2012} {[}Fig. \ref{fig:DynamicalDecoupling}(a){]}.
Fig. \ref{fig:DynamicalDecoupling}(b) shows the performance of dual-axis
XY4-N and XY8-N sequences, implemented using the time-optimal $\pi$-pulses
which were characterized above, compared to a single-axis sequence
implemented with the same pulses. A substantial suppression in the
pulse error accumulation is indicated by the slower decay coherence
as the number of $\pi$-pulses is increased. Furthermore, the inter-pulse
delay $\tau$ in these sequences was set to only two precession periods
of the spin, representing a regime of high pulse density which is
unreachable with weak harmonic driving techniques.

\section{Summary and outlook}

In this work we studied, both theoretically and experimentally, the
time-optimal construction of arbitrary single-qubit rotations under
a single strong driving field of finite amplitude. We showed that
arbitrary time-optimal rotations of two-level systems are constructed
as a series of bang pulses and drift periods, and that under a strong
driving field a combination of three such pulses can be used to obtain
general rotations on the Bloch sphere. This result is general to any
driven two-level system and may be applied to different physical realizations
such as quantum dots, donors in semiconductors, trapped ions and superconducting
flux qubits.

Importantly, our result allows for designing arbitrary \emph{rotations},
or single-qubit gates, rather than just steering the state of the
system from one point on the Bloch sphere to another, which is a crucial
requirement for implementations of quantum information processing
or quantum metrology. As an example, being able to apply $\pi$-rotations
around different axes on the Bloch sphere is a requisite for universal
dynamical decoupling sequences \cite{deLange2010}, as we demonstrated
in our study. This new regime of high-density pulse sequences, where
pulses are shorter than the qubit's precession period and their spacing
also approaches this timescale, can enable, for instance, quantum
sensing of high-frequency fields or efficient suppression of wide-band
decoherence processes.

Finally, in our study we used the dressed spin as a prototypical two-level
system with several key advantages: it benefits from longer coherence
times compared to the bare spin of the NV center, it is a more accurate
and robust realization of a true two-level system, and it offers better
controllability over the system parameters and the driving field.
Most importantly, it allows a straightforward and technically easy
approach to a strongly-driven quantum system, thus enabling fundamental
research of different manipulation techniques in this regime.
\begin{acknowledgments}
We thank Yoav Erlich and Konstanteen Kogan for technical assistance
with the experiments.
\end{acknowledgments}
\appendix*\section{Necessary conditions on time-optimal universal controls}

We provide here a complete proof that time-optimal control sequences,
for the problem presented in the main text, consist only of bang pulses
($\Omega=\pm\Omega_{\textrm{max}}$) and drift periods ($\Omega=0$).
A similar proof can be found in Ref. 19 of the main text. We begin
by parameterizing the time-evolution operator $U\left(t\right)$ using
three Euler angles,
\begin{equation}
U\left(\psi,\theta,\phi\right)=\exp\left(\frac{1}{2}i\psi\sigma_{z}\right)\exp\left(\frac{1}{2}i\theta\sigma_{y}\right)\exp\left(\frac{1}{2}i\phi\sigma_{z}\right),
\end{equation}
and define the state vector $\mathbf{x}\left(t\right)=\left(\psi\left(t\right),\theta\left(t\right),\phi\left(t\right)\right)$.
From Eqn. \ref{eq:GeneralHamiltonian} and \ref{eq:Schrodinger} we
find the equations of motion for this vector,
\begin{eqnarray}
\dot{\psi} & = & \omega_{1}-\Omega\cos\psi\cot\theta,\nonumber \\
\dot{\theta} & = & -\Omega\sin\psi,\label{eq:StateVectorEOM}\\
\dot{\phi} & = & \Omega\cos\psi\csc\theta.\nonumber 
\end{eqnarray}

We now apply Pontryagin\textquoteright{}s minimum principle (PMP)\cite{Pontryagin1987,Ross2009}
which gives the necessary conditions that a control $\Omega\left(t\right)$
must satisfy in order to be optimal, thus reducing substantially the
number of optimal control candidates. The principle states the following:
given a state vector $\mathbf{x}$ which satisfies a dynamical system
$\dot{\mathbf{x}}\left(t\right)=f\left(\mathbf{x}\left(t\right),\Omega\left(t\right);t\right)$,
where $\Omega\left(t\right)$ is a bounded control, we construct the
Pontryagin Hamiltonian
\begin{equation}
H_{P}\left(\mathbf{x}\left(t\right),\mathbf{p}\left(t\right),p_{0},\Omega\left(t\right);t\right)=\mathbf{p}\cdot\dot{\mathbf{x}}+p_{0}.
\end{equation}
Here $\mathbf{p}\left(t\right)$ is the costate vector, satisfying
the adjoint equation $\dot{\mathbf{p}}=-\partial H_{P}/\partial\mathbf{x}$
and $p_{0}$ is a non-negative constant chosen such that $\mathbf{p}\left(t\right)$
and $p_{0}$ do not vanish together at all times (\emph{non-triviality
condition}). According to the PMP, a necessary condition for a control
$\Omega\left(t\right)$ to be optimal is to maximize the Pontryagin
Hamiltonian $H_{P}\left(t\right)$ at all times. Additionally $H_{P}$
must vanish at the final time (\emph{transversality condition}). For
time-optimal problems in which $H_{P}$ has no explicit time-dependence,
it can be shown to be constant and equal to zero at all times under
optimal control \cite{Schattler2012}.

Using Eqs. \ref{eq:StateVectorEOM} we construct the Pontryagin Hamiltonian
for our problem,
\begin{eqnarray}
H_{P}\left(\mathbf{x},\mathbf{p},p_{0},\Omega\right) & = & -\Omega\left(p_{1}\cos\psi\cot\theta+p_{2}\sin\psi\right.\nonumber \\
 &  & \left.-p_{3}\cos\psi\csc\theta\right)+p_{1}\omega_{1}+p_{0},\label{eq:PMP_Hamiltonian}
\end{eqnarray}
with the costate vector $\mathbf{p}\left(t\right)=\left(p_{1}\left(t\right),p_{2}\left(t\right),p_{3}\left(t\right)\right)$.
We find that $H_{P}$ is linear in $\Omega$, and we define its coefficient
$\Phi\left(\mathbf{x},\mathbf{p};t\right)=\partial H_{P}/\partial\Omega$.
At times where $\Phi\left(t\right)$ is non-zero the Pontryagin Hamiltonian
is indeed a linear function of $\Omega$, and it obtains its maximum
at the edges of the allowed control range,
\begin{equation}
\Omega\left(t\right)=\begin{cases}
+\Omega_{\textrm{max}}\,\,\,\,\,\, & \Phi\left(t\right)>0,\\
-\Omega_{\textrm{max}}\,\,\,\,\,\,\,\, & \Phi\left(t\right)<0.
\end{cases}
\end{equation}

However when vanishes $\Phi\left(t\right)$ identically -- on a so-called
\emph{singular arc} -- $H_{P}$ becomes independent of $\Omega$ and
the maximization principle cannot be used to constrain it. Nevertheless
we now show, based on the non-triviality and transversality conditions,
that on a time interval where $\Phi\left(t\right)$ is identically
zero, the control $\Omega\left(t\right)$ must be zero as well. To
this end, we write explicitly $\Phi$ as
\begin{eqnarray}
\Phi\left(\mathbf{x},\mathbf{p};t\right)=\frac{\partial H_{P}}{\partial\Omega} & = & p_{1}\cos\psi\cot\theta+p_{2}\sin\psi\nonumber \\
 &  & -p_{3}\cos\psi\csc\theta.\label{eq:BigPhiDefinition}
\end{eqnarray}
We also write the equations for the costate vector $\mathbf{p}\left(t\right)=\left(p_{1}\left(t\right),p_{2}\left(t\right),p_{3}\left(t\right)\right)$
as derived from $\dot{\mathbf{p}}=-\partial H/\partial\mathbf{x}$,
\begin{eqnarray}
\dot{p}_{1} & = & \Omega\left(-p_{1}\sin\psi\cot\theta+p_{2}\cos\psi\right.\nonumber \\
 &  & \left.+p_{3}\sin\psi\csc\theta\right),\label{eq:P1_dot}\\
\dot{p}_{2} & = & \Omega\left(-p_{1}\cos\psi\csc^{2}\theta+p_{2}\sin\psi\right.\nonumber \\
 &  & \left.+p_{3}\cos\psi\csc\theta\cot\theta\right),\label{eq:P2_dot}\\
\dot{p}_{3} & = & 0.
\end{eqnarray}
We now set $\Phi\left(t\right)=0$ in Eq. \ref{eq:BigPhiDefinition},
and find that $p_{1}$ can be written as
\begin{equation}
p_{1}=-p_{2}\tan\psi\tan\theta+p_{3}\sec\theta.\label{eq:P1_singular}
\end{equation}
Also on a singular arc the Hamiltonian (\ref{eq:PMP_Hamiltonian})
has the form $H_{P}=p_{1}\omega_{1}+p_{0}$, and since it is an integral
of the problem, $p_{1}$ must be constant on singular arcs. Substituting
Eq. \ref{eq:P1_singular} into Eq. \ref{eq:P1_dot} and setting $\dot{p}_{1}=0$
we obtain
\begin{equation}
\Omega p_{2}\sec\psi=0,
\end{equation}
so that either $\Omega=0$ as required, or $p_{2}=0$. Assuming the
latter, we then must also have $\dot{p}_{2}=0$ on the singular arc,
and we find from substituting Eq. \ref{eq:P1_singular} into Eq. \ref{eq:P2_dot}
and setting $p_{2}=\dot{p}_{2}=0$,
\begin{equation}
-\Omega p_{3}\cos\psi\sec\theta=0.
\end{equation}
Again this means that either $\Omega=0$ as required, or that either
$\cos\psi$ or $p_{3}$ vanish identically on the singular arc. We
contradict the latter two possibilities:
\begin{enumerate}
\item If $\cos\psi=0$ on some time interval then $\psi$ must be constant
on that interval. This contradicts Eq. \ref{eq:StateVectorEOM} which,
for $\cos\psi=0$, has the form $\dot{\psi}=\omega_{1}\neq0$.
\item If $p_{3}=0$ then by Eq. \ref{eq:P1_singular} we have $p_{1}=0$,
since we also assumed $p_{2}=0$. This means that the costate vector
$\mathbf{p}$ vanishes entirely, and the Hamiltonian (\ref{eq:PMP_Hamiltonian})
is simply $H_{P}=p_{0}$. However based on the transversality condition,
$H_{P}$ must vanish on an optimal control, so $p_{0}=0$, and we
find that $\mathbf{p}=0$ and $p_{0}=0$ in contradiction to the non-triviality
condition.
\end{enumerate}
This completes the proof that $\Omega\left(t\right)=0$ on singular
arcs, i.e. for $\Phi\left(t\right)=0$.

\bibliographystyle{pnas}
\bibliography{StrongDriving_PRB}

\end{document}